\documentclass[10pt,conference]{IEEEtran}

\usepackage{color} 
\usepackage{hyperref} 

\pdfoutput=1

\usepackage{graphicx} 
\usepackage{listings}
\usepackage{booktabs}
\usepackage{tabularx}
\usepackage{array}
\usepackage{outlines}
\usepackage{boldline}
\usepackage{graphicx}
\graphicspath{{figs/}}
\usepackage{listings}
\usepackage{makecell}
\usepackage{mathtools}
\usepackage{multirow}
\usepackage{xspace}

\usepackage{xurl}

\usepackage{breakurl}
\usepackage{extarrows}
\usepackage{textcomp}
\usepackage{xparse}   
\usepackage{bigdelim}
\usepackage[linesnumbered,ruled,vlined]{algorithm2e}
\usepackage{color}
\usepackage{tikz}
\usepackage{pifont}
\usepackage{float}
\usepackage{svg}

\hypersetup{colorlinks=false, pdfborder={0 0 0}}

\begin{document}

\newcommand{\tpalit}[1]{\textcolor{magenta}{\textbf{Tapti: #1}}}
\newcommand{\alyssa}[1]{\textcolor{red}{\textbf{Alyssa: #1}}}
\newcommand{\gabe}[1]{\textcolor{orange}{\textbf{Gabe: #1}}}

\newcommand{\citeXXX}[1]{\textcolor{red}{[X]}\xspace}
\newcommand{\XXX}{\textcolor{red}{[XXX]}\xspace}
\newcommand{\YYY}{\textcolor{red}{[YYY]}\xspace}

\newcommand{\revision}[1]{\textcolor{blue}{\textbf{#1}}}

\newcommand{\sysname}{RustAssure\xspace}

\newcommand{\NUMMODELS}{four\xspace}
\newcommand{\NUMAPPS}{five\xspace}

\newcommand{\AVGCOMPILESUCCESSGPTFOUR}{{89.8\%}\xspace}
\newcommand{\AVGCOMPILESUCCESSGPTMINI}{{83.0\%}\xspace}
\newcommand{\AVGCOMPILESUCCESSGPTTHREEFIVE}{{70.5\%}\xspace}
\newcommand{\AVGCOMPILESUCCESSCLAUDE}{{48.3\%}\xspace}

\newcommand{\AVGUNSAFEGPTFOUR}{{13.05\%}\xspace}
\newcommand{\AVGUNSAFEGPTMINI}{{20.88\%}\xspace}
\newcommand{\AVGUNSAFEGPTTHREEFIVE}{{31.03\%}\xspace}
\newcommand{\AVGUNSAFECLAUDE}{{59.10\%}\xspace}

\newcommand{\AVGSCOREGPTFOUR}{{72\%}\xspace}
\newcommand{\AVGSCOREGPTMINI}{{74\%}\xspace}
\newcommand{\AVGSCOREGPTTHREEFIVE}{{71\%}\xspace}
\newcommand{\AVGSCORECLAUDE}{{80\%}\xspace}

\newcommand{\AVGCOVERAGELIBCSV}{{94.6\%}\xspace}
\newcommand{\AVGCOVERAGEOPTIPNG}{{82\%}\xspace}
\newcommand{\AVGCOVERAGEURLPARSER}{{79.2\%}\xspace}
\newcommand{\AVGSCOVERAGELIBBMP}{{83.6\%}\xspace}
\newcommand{\AVGSCOVERAGEUC}{{95\%}\xspace}

\newcommand{\AVGCOVERAGELIBCSVRUST}{{90.85\%}\xspace}
\newcommand{\AVGCOVERAGEOPTIPNGRUST}{{78.97\%}\xspace}
\newcommand{\AVGCOVERAGEURLPARSERUST}{{78.59\%}\xspace}
\newcommand{\AVGSCOVERAGELIBBMPRUST}{{83.53\%}\xspace}
\newcommand{\AVGSCOVERAGEUCRUST}{{92.55\%}\xspace}

\newcommand{\SYMBOLICEXECTIMEOUT}{{180 minutes}\xspace}
\newcommand{\AVGCOVERAGE}{{62.9\%}\xspace}

\newcommand{\purge}[1]{{\textbf{#1}}}
\renewcommand{\purge}[1]{}
\newcommand{\APPS}{9\xspace}
\newcommand{\AVGOVERHEAD}{5.45}
\newcommand{\AVGPTSIMP}{13.15}
\newcommand{\MAXPTSIMP}{1.25}
\newcommand{\PYLOC}{{139}\xspace}
\newcommand{\SYMEXECTIMEOUT}{\textcolor{blue}{\textbf{10800}}\xspace}

\renewcommand{\tpalit}[1]{}
\renewcommand{\gabe}[1]{}
\renewcommand{\citeXXX}{}
\renewcommand{\XXX}{}
\renewcommand{\YYY}{}
\renewcommand{\revision}{}

\newcommand{\intel}{Intel\textsuperscript{\textregistered}\xspace}
\newcommand{\intels}{Intel's\textsuperscript{\textregistered}\xspace}

\newcommand{\critarg}{Precision Critical Arguments\xspace}
\newcommand{\LIH}{optimistic analysis\xspace}
\newcommand{\LINH}{baseline analysis\xspace}
\newcommand{\LIHMV}{optimistic MV\xspace}
\newcommand{\LINHMV}{fallback MV\xspace}
\newcommand{\Addrof}{\emph{Addr-Of}\xspace}
\newcommand{\Deref}{\emph{Load}\xspace}
\newcommand{\Store}{\emph{Store}\xspace}
\newcommand{\Copy}{\emph{Copy}\xspace}
\newcommand{\FieldOf}{\emph{Field-Of}\xspace}

\def\Snospace~{\S{}}
\def\sectionautorefname{\Snospace}
\def\subsectionautorefname{\Snospace}
\def\subsubsectionautorefname{\Snospace}


\newcommand\para[1]{\vspace{0.4em} \noindent \textbf{#1}}
\newcommand\parasmall[1]{\vspace{1em} \noindent \textbf{#1}}

\newcommand{\squishenumerate}{
   \begin{enumerate}
       { \setlength{\itemsep}{0pt}  \setlength{\parsep}{3pt}
      \setlength{\topsep}{3pt}       \setlength{\partopsep}{0pt}
      \setlength{\leftmargin}{2em} \setlength{\labelwidth}{1em}
      \setlength{\labelsep}{0.5em} } }

\newcommand{\squishenumerateend}{
    \end{enumerate}  }

\newcommand{\squishlist}{
   \begin{list}{$\bullet$}
    { \setlength{\itemsep}{0pt}      \setlength{\parsep}{3pt}
      \setlength{\topsep}{2pt}       \setlength{\partopsep}{0pt}
      \setlength{\leftmargin}{2em} \setlength{\labelwidth}{1em}
      \setlength{\labelsep}{0.5em} } }

\newcommand{\squishlisttwo}{
   \begin{list}{$\bullet$}
    { \setlength{\itemsep}{0pt}    \setlength{\parsep}{0pt}
      \setlength{\topsep}{0pt}     \setlength{\partopsep}{0pt}
      \setlength{\leftmargin}{2em} \setlength{\labelwidth}{1.5em}
      \setlength{\labelsep}{0.5em} } }

\newcommand{\squishend}{
    \end{list}  }




\title{\sysname: Differential Symbolic Testing for  
LLM-Transpiled C-to-Rust Code}


\author{\IEEEauthorblockN{1\textsuperscript{st} Yubo Bai}
\IEEEauthorblockA{\textit{Department of Computer Science} \\
\textit{University of California, Davis}\\
Davis, USA \\
gabbai@ucdavis.edu}
\and
\IEEEauthorblockN{2\textsuperscript{nd} Tapti Palit}
\IEEEauthorblockA{\textit{Department of Computer Science} \\
\textit{University of California, Davis}\\
Davis, USA \\
tpalit@ucdavis.edu}
}

\maketitle


\begin{abstract}
Rust is a memory-safe programming
language that significantly improves
software security. Existing codebases written in unsafe memory languages, such as C, must
first be
transpiled to Rust to take advantage of Rust's improved safety guarantees. 
\sysname presents a system that uses Large Language Models (LLMs) to
automatically transpile existing C codebases to Rust. \sysname uses 
prompt engineering techniques to maximize
the chances of the LLM generating \emph{idiomatic} and 
\emph{safe} Rust code. Moreover, because LLMs often
generate code with subtle bugs that can be missed
under traditional unit or fuzz testing, 
\sysname performs differential symbolic testing
to establish the semantic similarity
between the original C and LLM-transpiled Rust code. 
We evaluated \sysname
with \NUMAPPS real-world 
applications and libraries, and showed that 
our system is able to generate compilable Rust 
functions for 89.8\%
of all C functions, of which 72\% produced equivalent symbolic
return values for both the C and Rust functions.
\end{abstract}

\section{Introduction}

Rust is a memory-safe programming language that is 
increasingly gaining popularity.
Automatically transpiling codebases
written in memory-unsafe languages 
such as C, to Rust, can significantly improve software
security. 
Previous syntax-directed and program analysis-driven methods~\cite{emre2021translating, emre2023aliasing, 
zhang2023ownership}
have achieved limited success,
correctly transpiling only a fraction of the
C code to safe Rust.
Moreover, the Rust code generated by these approaches 
is
difficult to read and 
often does not follow idiomatic Rust conventions~\cite{eniser2024towards}.
More recently, Large language models (LLMs)
have shown significant promise in code 
completion~\cite{copilot}, code generation~\cite{vaithilingam2022expectation, dong2024self,
mu2024clarifygpt}, and code transpilation~\cite{eniser2024towards}.
However, LLMs are not guaranteed to always
generate correct code~\cite{pan2024lost}. Therefore, it is essential to 
verify the functional correctness of LLM-generated code.

Existing approaches for testing LLM-transpiled Rust
code suffer from various limitations.
Approaches based on unit testing
inherently
suffer from code coverage issues, especially for 
large complex applications, leading to 
significant portions of the code remaining untested. 
Approaches based on differential 
fuzzing~\cite{eniser2024towards} 
compare the 
test case outputs for the original and transpiled code.
However, the exact return values of the C and Rust functions
can differ
due to runtime differences. 
For example, the values of 
pointers returned by the functions
would differ across different language runtime environments. Therefore, a mismatch in the concrete runtime values
do not necessarily indicate 
an error. 

To mitigate these challenges, we present \sysname---a system that 
uses Large Language Models (LLMs) to transpile
C code into Rust and establishes the \emph{semantic similarity} between
the original and transpiled code using \emph{differential symbolic
testing.} Unlike dynamic testing approaches such as unit or fuzz testing, where the transpiled Rust function 
is executed with concrete inputs, \sysname \emph{symbolically} executes both the original
C and LLM-transpiled Rust code to detect 
functional
divergence between the original C and Rust code. Unlike dynamic testing
which is \emph{input-specific}, 
symbolic execution can explore all paths, thus 
proving that all inputs produce equivalent
outputs and guaranteeing correctness, while differential testing can only
show that the behavior is the same for the executed input. Furthermore, 
by detecting the divergence at the symbolic level, \sysname
avoids false positives stemming from differences in the runtime environment, such
as memory layout and heap addresses.

\sysname first transpiles the C code to Rust using Large Language Models (LLMs).
Multiple factors must be considered when using LLMs to transpile C code to Rust. First, the LLM-based transpiler must ensure
that the C code is transpiled to \emph{safe} Rust to the greatest
extent possible. To provide programmers
with more flexibility, Rust 
provides an ``escape-hatch'' that relaxes 
certain memory safety rules if the code is enclosed in 
an \texttt{unsafe} block. Unsafe
Rust code introduces potential memory corruption vulnerabilities in the program, therefore, the amount 
of unsafe Rust code generated by the LLM must be minimized.
\sysname uses various prompt-engineering
techniques to ensure that the 
code generated by the LLM compiles successfully,
and also minimizes the use of unsafe Rust. 

After the LLM generates compilable Rust code, 
\sysname performs differential symbolic testing 
on the original C and LLM-transpiled Rust code. 
\sysname contains a \emph{Semantic 
Similarity Checker} that symbolically executes individual
C functions and their LLM-transpiled Rust counterparts, deriving symbolic values for all implicit and explicit return 
variables of the function. \sysname then compares the symbolic value for each return value of the function
and reports a \emph{Semantic Similarity Score (S$^3$
score)} for the Rust function. 
Moreover, differential symbolic testing can not
only identify which return values diverge, but also
\emph{how} they diverge, making it easier for the
programmer to pinpoint the bug.

\sysname's symbolic execution is \emph{composable.} 
\sysname can analyze each Rust function individually and determine if it is semantically equivalent to the 
original C function. If the Rust function being analyzed 
invokes another function, \sysname abstracts away the implementation of that function and only ensures that the
called function returns symbolic values. This allows the caller function to be symbolically tested even 
if the called function does not compile or link successfully with the rest of the LLM-transpiled 
Rust code. As a result, the programmer can test each LLM-transpiled 
Rust function for semantic equivalence and improve their 
confidence in the LLM-produced code, even if a few LLM-generated Rust functions
do not successfully compile. In fact, from our experience even with state-of-the-art
LLM models such as GPT-4o~\cite{gpt-models}, only 89.8\%
of all transpiled
C functions successfully compile without user intervention. \sysname is particularly useful in such scenarios.

Establishing semantic equivalence using symbolic return values across different languages is not
trivial. Often, Rust code uses types that are missing in C completely. For example,
safe Rust code does not allow explicit \texttt{NULL} values but requires \emph{wrapping} values
that can potentially be \texttt{NULL} in an \texttt{Option} \emph{enum} type. 
Accurate
extraction and comparison of symbolic values across different languages requires \emph{aligning}
these differing C 
and Rust
types correctly. \sysname identifies three 
such commonly occurring patterns and maps the Rust
type correctly to the C type, thus facilitating the 
symbolic value comparison of the return values. 


We used \sysname to transpile five C real-world codebases to Rust, and used the \emph{Semantic Similarity
Checker} to compute the \emph{Semantic Similarity Score 
(S$^3$ score)} for the transpilations.
These libraries had a total of 176 functions,
of which 158, 146, 124, and
85 functions were compiled successfully 
by the GPT-4o, GPT-4o-mini, 
GPT-3.5-turbo~\cite{gpt-models, gpt-4-o-mini} and 
Claude-3.5-Sonnet~\cite{claude} models, respectively. 
Across all of these LLM models, \sysname's \emph{S$^3$} metric 
allowed us to find 12 complex
semantic bugs, along with 13 
simple bugs. Because these 12 and 13 bugs
are present in more than one LLM model, the total number of 
buggy functions detected are 20 and 17, for semantic and simple
bugs, respectively.
Moreover, 
we computed the average precision of the S$^3$ score, 
across all LLM models
to be 85.7\% and 88.2\%, for the \emph{libcsv} and \emph{u8c} codebases, 
respectively, while the average recall was 100\% for both codebases. 
An end-to-end comparison shows that \sysname
can generate code with 72\% fewer raw pointers than
the state-of-the-art syntax-directed C-to-Rust transpilation
toolchain, CROWN~\cite{hanliangownership}, and find 11
semantic bugs
that the testing framework, 
FLUORINE~\cite{eniser2024towards}, is
unable to find.
In summary, we make the following contributions---

\squishlist
    \item \emph{\sysname}, an LLM-based pipeline for transpiling C code to safe 
    Rust. The pipeline uses various prompt engineering
    techniques
    to assist the LLM in generating safe Rust code.
    \item \emph{Semantic Similarity Checker}, a symbolic execution-based system to identify the semantic similarity between the original C and LLM-transpiled Rust code.
    \item Three patterns for reconciling the type differences between C and Rust that make it possible to compare the
    symbolic return values in C and Rust code.
    \item Evaluation of \sysname using five real-world
    libraries that demonstrate \sysname's bug-finding
    capabilities, and comparisons against the
    CROWN~\cite{hanliangownership} and FLUORINE~\cite{eniser2024towards} C-to-Rust toolchains. 

\squishend
Our 
toolchain can be found at 
\url{https://github.com/davsec-lab/rustassure}.

\section{Motivating Example}
\label{sec:u8_next}
\begin{figure*}
    \centering
    \includegraphics[width=\textwidth]{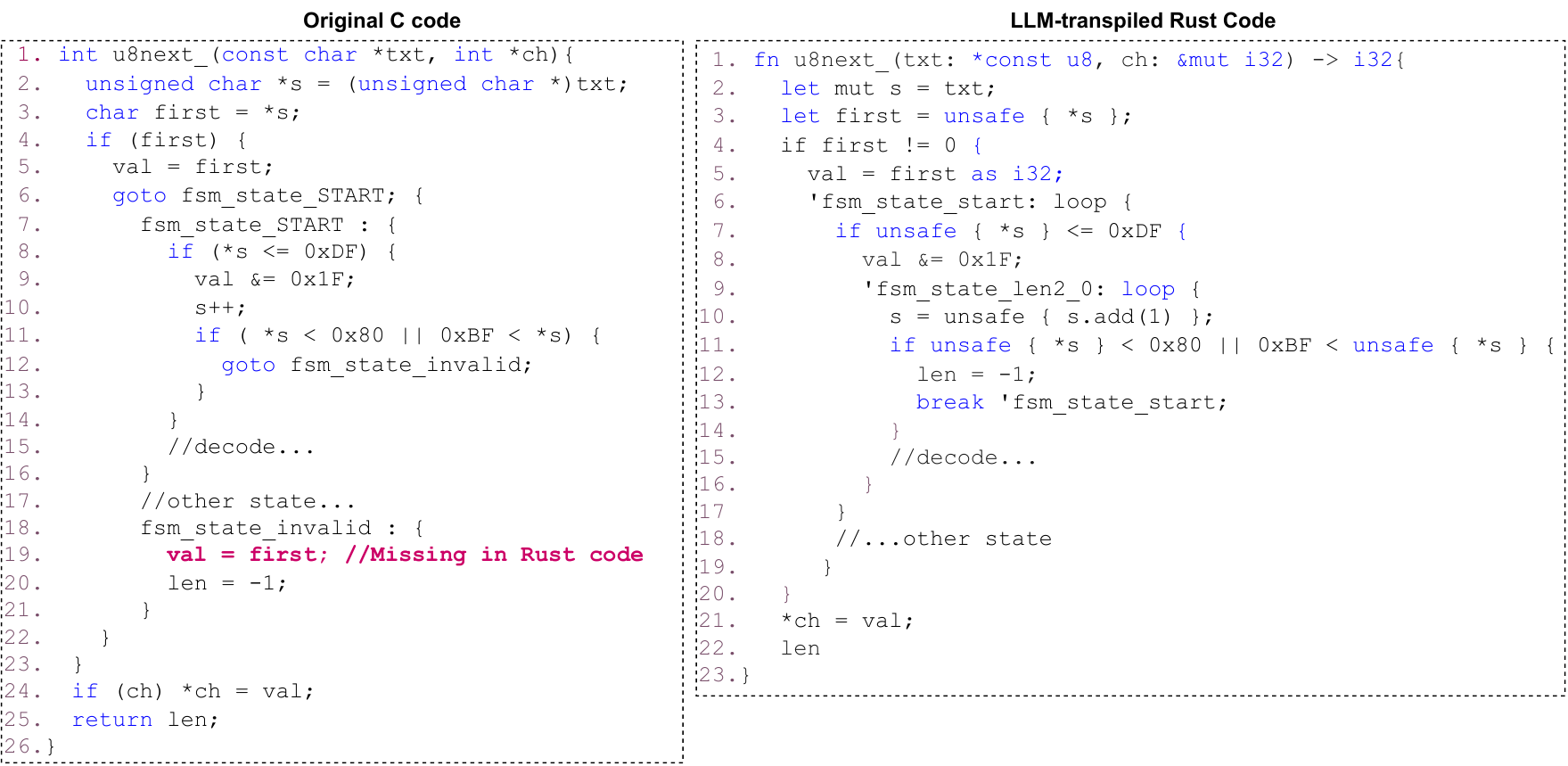} 
    \caption{Bug detected by \sysname. When executed
    with an invalid UTF string, the LLM-generated
    Rust code returns an incorrect intermediate state instead of
    a pointer to the start of the input string, which is the expected
    functionality.}
    \label{fig:motivating_example}
\end{figure*}



Code generated by Large Language Models (LLMs) often 
contain subtle bugs.
Consider the simplified code snippet for the function \texttt{u8next\_} 
from the \emph{u8c} library,
shown in 
\autoref{fig:motivating_example}. 
The original C function takes as arguments
an input
UTF-8-encoded string, \texttt{txt}, and an 
output argument,
\texttt{ch}.
It returns the number of bytes encoding the
first Unicode codepoint of the string \texttt{txt}, and stores 
the converted Unicode codepoint in \texttt{ch}.
If the encoding is not valid, it
returns \texttt{-1} 
and 
stores the \emph{first byte of the input string} in 
\texttt{ch}.

To perform the conversion, the function maintains a state machine,
as shown in lines 6-23 in the original 
C code.
The LLM-transpiled Rust code contains a subtle
bug in this state machine code.
When the input is not a valid UTF string 
and contains
an invalid byte in any of the \emph{non-first} byte offsets,
the functionalities of the original C and the 
LLM-transpiled Rust functions diverge.
For example, consider the 
two character input sequence \texttt{\{0xC2, 0x41\}}. 
The first byte, \texttt{0xC2} is valid, but 
the second byte, \texttt{0x41}, is invalid.
When executed with this byte sequence, 
the C program will return the original 
first byte, which is \texttt{0xC2}, but the transpiled 
Rust program will return an intermediate decoding state,
specifically, the bit-masked value \texttt{0x02}. 
This occurs because the generated Rust code is missing
the operation that resets the variable \texttt{val}
to the first character of the input string \texttt{txt} 
(line 19, in the C code). The variable \texttt{val} is finally copied
to the \texttt{ch} argument, resulting in the functional
divergence.
Thus, the C and Rust API semantically
diverge, and if the users of the Rust API 
are not aware of this change 
they may misinterpret error conditions or incorrectly
assume 
behavior consistent with the C API.

Detecting this bug requires knowledge of Unicode/UTF-8 semantics, 
which traditional fuzzers lack,
limiting their effectiveness.
Moreover, the unit
test cases provided for the \emph{u8c} library 
do not verify the returned byte
in the \texttt{ch} output argument
when an invalid UTF string is provided, and only
focuses on the explicit return value for checking error
conditions. 
Because the \sysname toolchain relies on symbolic execution, 
it can correctly identify this semantic divergence.
\section{Design}

\begin{figure*}
    \centering
    \includegraphics[width=0.9\textwidth]{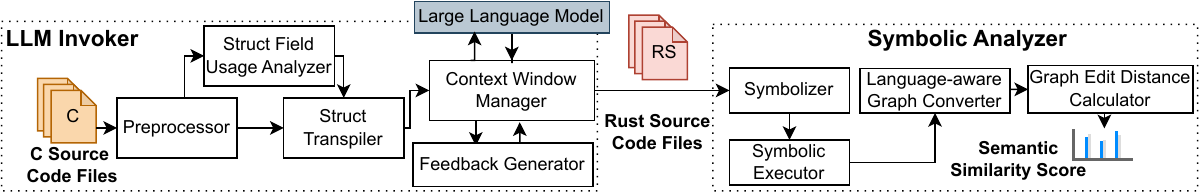} 
    \caption{\sysname transpiler toolchain.}
    \label{fig:blockdiag}
\end{figure*}

\autoref{fig:blockdiag} shows \sysname's two major components---the LLM
Invoker and the Symbolic Analyzer. 
The LLM Invoker is responsible
for invoking the LLM model for the 
transpilation task, followed by the Symbolic Analyzer
performing differential symbolic testing to identify
functional divergences.

\subsection{LLM Invoker}
\label{sec:llminvoker}

\sysname's LLM Invoker is responsible for taking a C codebase,
preprocessing it, and invoking the configured LLM model 
for transpilation. 

\para{Header File Preprocessor.} 
We observed that naively 
sending the code in the C source code files (\texttt{.c} files)
to the Large
Language Model for transpilation is not adequate. 
This code often lacks much of the contextual information that is
essential for the correct transpilation of each function 
in the source code file, such as,  
\texttt{struct}
definitions, function declarations, preprocessor
macros, and \texttt{typedef} aliases,
which are defined in 
\texttt{.h} header files.

However, when the C preprocessor processes the \texttt{.c} file,
it naively adds all contents of all 
header files included by the \texttt{.c} file, even 
if the content is not actually used by the C code.
Therefore, including all output produced by the C preprocessor
in the LLM request
is unnecessary. Moreover, LLM models have a fixed 
context window~\cite{context-window} that limits the amount
of input it can process at a time. Therefore, \sysname's
LLM Invoker includes only the \emph{essential} macro expansions,
\texttt{typedef} instructions, and \texttt{struct} and 
\texttt{function} definitions in the LLM request.
This contextual information helps the LLM produce accurate
transpilations.


\para{Struct Transpiler.}
\label{design:replay}
Multiple functions, whose transpilation spreads out over multiple
LLM invocations can use the same \texttt{struct} type. We
observed that due
to the inherent non-determinism and randomness~\cite{temperature}
of LLM models
these different invocations can sometimes generate slightly
different Rust \texttt{struct} definitions for the same 
C types. For example, the names of certain transpiled \texttt{struct} fields
can differ between different LLM invocations. These
differing definitions cause linkage issues when the
transpiled Rust functions are linked together to build the 
final executable.
To avoid such linkage issues, the LLM Invoker first
extracts all C \texttt{struct} definitions and invokes the LLM
to \emph{pre-translate} them. These Rust transpilations of 
\texttt{struct} types are stored in a \emph{Struct Transpilation Cache}.
Then, for each function that uses
these \texttt{struct} types, the pre-generated \texttt{struct}
definition is provided along with the C function as the prompt,
and the LLM is instructed to use the pre-generated definition.
This ensures consistent transpilation of \texttt{struct}
types across different functions.

\para{Struct Field Usage Analyzer.}
In many cases, the correct Rust type for a variable
depends on the \emph{usage} of the corresponding C variable.
For example, in C/C++, \texttt{struct} fields of \texttt{char*} type
can be used to store
both null-terminated strings, or arrays of characters. The
usage of the \texttt{char*} field must be analyzed to determine
whether it is used as a string
or an array. In fact, without such analysis, from our experience,
LLM models
transpile such \texttt{struct} 
fields into Rust \emph{raw pointers}, such 
as \texttt{*mut T}, and enclose any code using these pointers
fields
with \texttt{unsafe} blocks.
To limit the amount of unsafe Rust code, the LLM Invoker extracts
all uses of such \texttt{char*} fields and adds them to the LLM
request, when pre-translating 
the \texttt{struct} types, and instructs the LLM to consider these field usages
for the transpilation task.


\para{Context Window Manager.}
While LLM context window sizes are ever increasing, 
many LLM models, such as GPT-3.5-turbo~\cite{gpt-models} and Mistral 7B~\cite{mistral}
have smaller context windows of 16K tokens. 
When using these models,
a single large function can 
exceed the context window length.
To support these
models, the LLM Invoker automatically 
\emph{chunks} a function into
multiple LLM requests if it does not fit within
the LLM's context window. The LLM's 
response for each chunk is then collected
and assembled into the final Rust function.

\para{Feedback Generator.}
Like previous approaches~\cite{deligiannis2023fixing},
when the LLM produces a transpilation that does not
compile, \sysname re-attempts the transpilation,
by sending the compiler error message along with the original C 
function to the LLM, requesting a corrected translation.
This approach leverages the precise and diagnostically
rich error messages emitted by the Rust compiler to 
provide feedback to the LLM, thus increasing the chances
of it fixing the offending code.
In addition to compilation errors, if the LLM-transpiled Rust
code contains any \texttt{unsafe} block, or invokes unsafe
functions such as the functions from the 
\texttt{libc} crate, the LLM Invoker requests a re-translation
up to a configurable number of times. Thus,
each C function is transpiled using the LLM, maximizing the
possibility of safe Rust code.

\subsection{Symbolic Analyzer}
After the LLM Invoker produces compilable Rust code,
\sysname symbolically
executes them to derive the symbolic return values for 
each function. We note that C and Rust functions
will not be symbolically equivalent
if the C code has
memory corruption bugs that are absent in Rust---a limitation that is present
even when using concrete test cases. Such cases can result in false positives.
\tpalit{We can move this elsewhere, but it shouldn't be an
entire Discussion section.}

\para{Symbolizer.}
The Symbolizer instruments the C and Rust
counterparts of each
function with a \emph{prologue} and an \emph{epilogue.} 
The function prologue
initializes the function arguments with symbolic values, 
enabling the Symbolic Executor to 
symbolically execute the function under test.
If the function takes pointer arguments or
arguments of \texttt{struct} type, the Symbolizer ensures that 
symbolic memory regions are allocated for each pointer and
\texttt{struct} field. Pointer type arguments and fields can
be nested to arbitrary levels. For example, an argument
of \texttt{struct} type can contain a pointer field pointing to 
another \texttt{struct} object. The Symbolizer symbolizes
nested pointers and nested \texttt{struct} fields up to a 
configurable nesting limit. For our evaluation, we use
a maximum limit of 10.

The Symbolizer also instruments the function execution completion
with a \emph{function epilogue}
to log the symbolic values of the explicit return 
value and output parameters.
For complex types such as \texttt{struct} and array types, 
the Symbolizer logs the symbolic values for each field or element.
If these fields or elements are pointers, the Symbolizer recursively
traverses the symbolic object pointed to by its pointer, and logs
its symbolic values.
The Symbolizer tracks which fields of 
\texttt{struct} type output parameters were modified and logs the
symbolic values of only those fields.
This selective logging ensures that only meaningful symbolic state changes are reported.

The Symbolizer also instruments any external function calls to return 
symbolic values, ensuring that symbolic execution can proceed 
without requiring concrete results from unknown or unanalyzed 
functions. By replacing the outputs of external function 
with symbolic 
expressions, our analysis remains fully composable, and each
function can be symbolically executed independently.

\para{Symbolic Executor.}
After the Symbolizer modifies the C and Rust functions, the
Symbolic Executor performs the core symbolic execution process
using the KLEE~\cite{cadar2008klee} symbolic execution engine. When the
symbolic execution completes, the \emph{function epilogue}
inserted by the Symbolizer is executed, and the symbolic values
of the function return values are printed.
To facilitate
the symbolic execution of both C and Rust code using KLEE, 
we convert both the
original C and the transpiled Rust code to an Intermediate
Representation (IR) and symbolically execute this IR code. 
However, symbolically executing
the Rust-produced IR code adds Rust language-specific
artifacts to the symbolic return values. We discuss how 
we handle these language-specific artifacts in the symbolic
values in \autoref{design:graphconv}.


\subsection{Semantic Similarity Checker}
The \emph{Semantic Similarity Checker} compares the 
semantic equivalence between the original C code and the 
LLM-transpiled Rust code by comparing the symbolic return
values of the corresponding C and Rust functions. The
symbolic values are represented by a Backus-Naur Form grammar, called
\emph{KQuery}~\cite{kquery}. \sysname processes these KQuery symbolic
values as discussed below.

\para{Language-aware Graph Converter.}
\label{design:graphconv}
\begin{figure}
    \centering
    \includegraphics[width=\columnwidth]{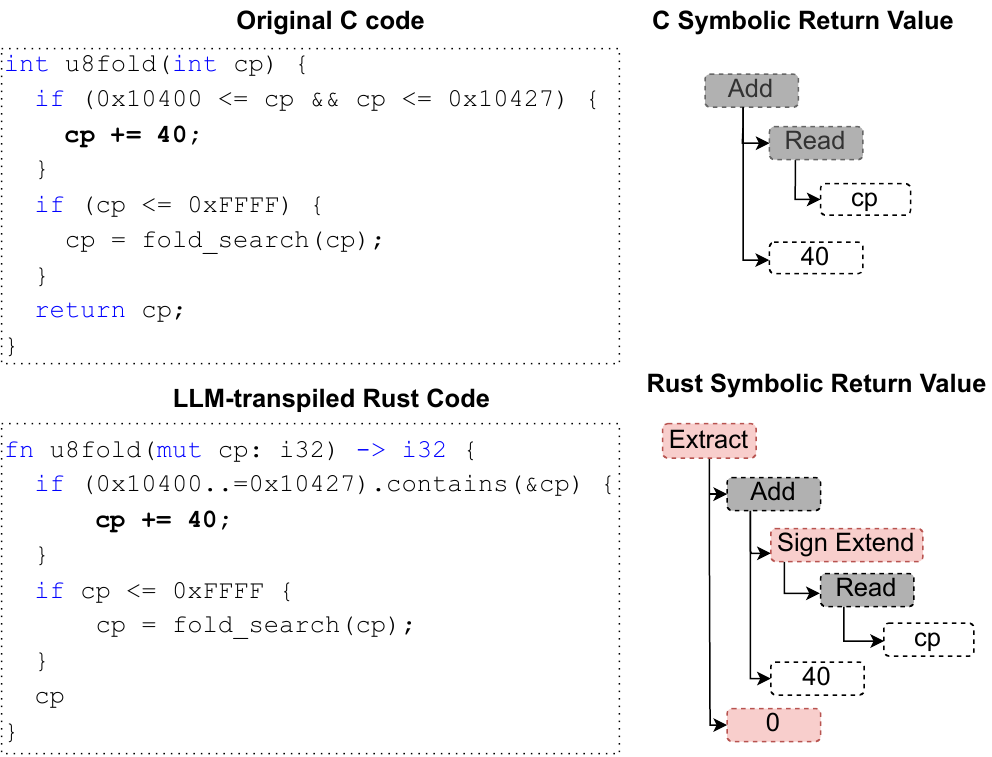} 
    \caption{C and Rust symbolic return values for the \texttt{u8fold}
    function. The red boxes
    are language-specific artifacts added by the Rust compiler. }
    \label{fig:graphnormalization}
\end{figure}
The \emph{Language-aware Graph Converter} component converts the symbolic values
to a graphical representation. 
In some cases, language differences between C and Rust can 
cause differences in symbolic representations between C and
Rust, even when the high-level semantics of the code remain the same.
The graph representations allows us
to easily apply normalization techniques on the symbolic values due to 
such language-level differences.

Consider the C and transpiled Rust code, from the \emph{u8c}
library, shown in 
\autoref{fig:graphnormalization}. The function \texttt{u8fold} maps a 
Unicode codepoint to its lowercase form. 
Although both the C and Rust code perform the same \texttt{cp += 40}
operation in the first \emph{if} block, 
the symbolic output
graphs differ due to Rust’s explicit type extension operations needed to 
support integer overflow checks necessary for memory safety. These checks are completely missing in C.
These additional nodes in Rust’s symbolic execution graph cause
a mismatch even when the underlying operation is 
identical.
To address these differences, the Language-aware Graph Converter 
applies graph normalization techniques to remove redundant type extension operations that do not affect program semantics, by
removing their corresponding nodes from the symbol graph.
    

\para{Graph Edit-distance Calculator.}
\sysname uses the graph-edit distance between the
corresponding graphical
representations of the symbolic return values of the C and
Rust functions as the Semantic Similarity Score \emph{ ($S^3$ score)}
between return values in
the original C code and the LLM-transpiled Rust code.
An edit distance of zero means that the Rust function
returns the same symbolic return value as the original C function.
A higher edit distance implies a higher degree of
semantic divergence.
\sysname uses a standard graph edit-distance algorithm~\cite{abu2015exact}
to compute the edit distance between the return values. We 
used a graph edit-distance-based comparison instead of SMT solving to 
simplify our implementation. In the future, we plan to explore
using SMT solvers to establish strict equivalence.
\tpalit{Gabe: I removed a lot of the text you had added. We don't need to explain in so much detail here. We're still in the design section,
we can't get into so much details.}


\tpalit{Commenting out this part as it is only a runtime
optimization.}

\section{Handling Language-level Differences}
\label{sec:bridging}

Fundamental differences between C and Rust
often results in a mismatch between types and 
object memory layout 
when variables
are
transpiled from C to Rust. To limit
false positives, \sysname aligns these mismatches as described
below.


\subsection{Aligning Memory Layout.}
\label{sec:bridging:mem-layout}
We observe that the memory
layouts of complex types such as \texttt{structs}
generated by C compilers, such as Clang~\cite{clang},
differ from the memory layouts generated by the Rust compiler,
\texttt{rustc}.
By default, the Rust compiler aligns fields differently 
and adds additional padding in between different fields. 
Moreover, the Rust compiler often also
reorders the struct fields for 
memory optimization. This differing memory layout
makes it challenging 
to compare the symbolic value of the
original C objects
and the transpiled Rust objects. 

Fortunately, Rust allows the programmer to specify alternative data 
layout strategies. In particular, it
provides the attribute \texttt{repr(C)} to specify 
a C-like memory layout. Similarly, the 
\texttt{repr(packed)} attribute instructs the compiler
not to insert any additional padding between the \texttt{struct} fields. 
Therefore, \sysname applies the \texttt{\#[repr(C, 
packed)]} attribute on every Rust struct type. This ensures
that the
compiler lays the struct fields out in memory exactly the way a C 
compiler would, keeping struct fields in order and using C-style alignment 
and padding. This simplifies the recovery of the symbolic return 
values from these fields.

\subsection{The \emph{Option} type.}
\label{sec:bridging:optiontype}
Unlike C, Rust does not support \texttt{NULL} values
and instead uses the \texttt{Option<T>} enum to 
represent an \emph{optional} value.
The \texttt{Option<T>} enum 
type can either contain a value of type \texttt{T (Some(T))} 
or be empty \texttt{(None)}. 
Thus, C pointer variables that can potentially be
\texttt{NULL}, are wrapped in the \texttt{Option<T>} enum
type
in Rust.
Therefore, the apparent type of the C and transpiled Rust variable
will not be the same. This difference means that
\sysname cannot simply use the return value types
to retrieve and compare the 
symbolic values of the C and Rust variables.
Additional processing is required to align the
C and Rust types in such cases.


However, in spite of the difference in the types of the
variables, when the Rust compiler lowers the \texttt{Option}
enum type variable to the machine code, it treats the 
enum variable simply as \emph{syntactic sugar}
around the inner object. The memory layout 
of the original C nullable pointer of type \texttt{T} 
and Rust's 
\texttt{Option} wrapper around the type \texttt{T}
are the same, assuming the requirements outlined in 
\autoref{sec:bridging:mem-layout} are followed. 
The only special operations the Rust compiler
inserts are \texttt{NULL} checks that panic 
on failure.
Therefore, when \sysname's Semantic Similarity Checker
encounters a situation
where a return value or output parameter in Rust is of 
\texttt{Option} type, it simply casts it to the 
underlying type
to access its symbolic representation.

\subsection{String and Slice types.}

The C language represents strings as null-terminated arrays of \texttt{char} type. 
The corresponding type in Rust is \emph{String}.
A related Rust type is the slice type \emph{\&str} that 
represents an immutable reference to a sequence of characters. Similar to string slices, Rust also
allows slices of arrays.
Both \texttt{String} and slice type variables are 
internally represented by 
\texttt{struct} types by the Rust compiler.
The first field of this \texttt{struct} is a \emph{data} pointer pointing to 
the array of \texttt{char} type, and the second 
field represents
the length. Therefore, for return values of these types,
\sysname's Semantic Similarity Checker accesses the first field of the \texttt{struct} type and casts it 
to \texttt{char*} before comparing its symbolic values. 

Internally, the Rust compiler initializes the slice types,
\texttt{\&str} and \texttt{\&[T]} as arrays of initial
length zero, indicating that their actual length is determined
at runtime. When symbolizing such slice types, \sysname's
Symbolizer
initializes these slices to symbolic buffers of size \texttt{100}
and
explicitly initializes the \texttt{len} field of 
the struct to \texttt{100} to 
safely access its contents.


\subsection{The \emph{vector} type.}
C arrays are transpiled to Rust's \texttt{Vec} type objects.
Unlike C arrays,
Rust guarantees that any out-of-bounds accesses
in a \texttt{Vec} type will 
cause a panic at runtime.
Similar to the \texttt{String} type, the \texttt{Vec<T>} type is represented by a \texttt{struct} type that contains 
three fields,
a \emph{data} pointer to the array of type \texttt{T}, 
a \emph{len} field to represent the length of the array
indicating the actual number of elements contained,
and a \emph{capacity} field to represent the capacity
of the array. The semantic checker directly accesses this field and compares
the symbolic values of each array element in the Rust code with its C counterpart.

\begin{figure}
    \centering
    \includegraphics[width=\columnwidth]{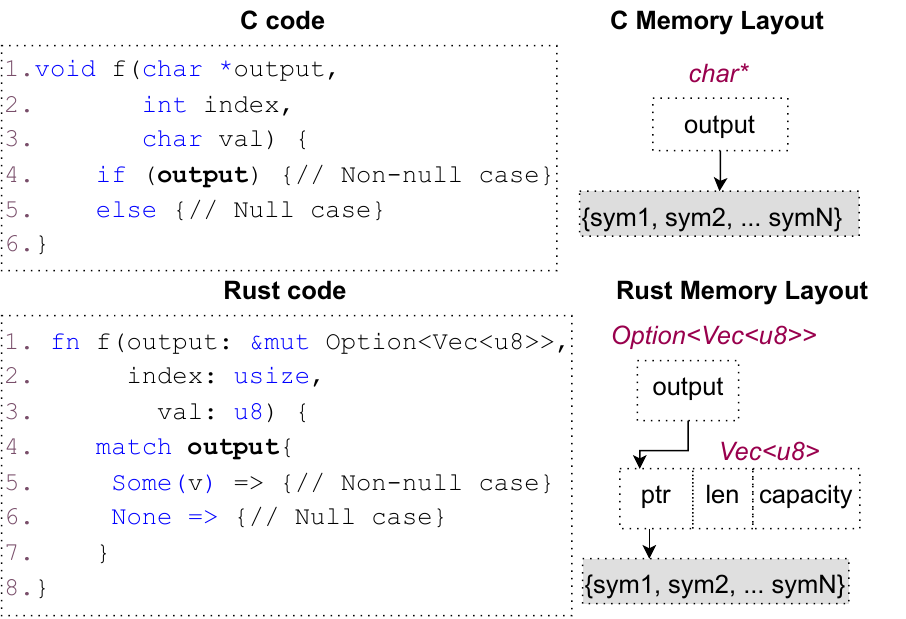} 
    \caption{Differing memory layout between C and Rust.
    The argument \texttt{input} is a \texttt{char*} array
    for which the transpiled Rust code uses \texttt{Vec}
    type.}
    \label{fig:option-example}
    \vspace{-1em}
\end{figure}
%
When the \texttt{String}, \texttt{str\&}, \texttt{Vec} are 
wrapped in an \texttt{Option}, 
we extract and cast the inner value as explained in 
\autoref{sec:bridging:optiontype}, before comparing the 
symbolic values.
\autoref{fig:option-example} shows an example 
of a nullable array of characters, \texttt{output},
being transpiled 
into a Rust \texttt{Vec} object, wrapped in an 
\texttt{Option} enum object. The figure also 
shows the C and Rust
code used to access the respective objects, along with
the objects' memory layouts. 
As shown, the Rust variable \texttt{output} of type 
\texttt{Option<Vec<u8>>} is a pointer
to the \texttt{Vec} object that contains three fields, of
which the first \texttt{ptr} field points to the actual data buffer.
The actual memory layout of the data buffer referenced by
the \texttt{ptr} field of Rust's \texttt{Vec} object
is the same as
the layout of the C \texttt{output} variable. 
Therefore, for the Rust code, \sysname's Semantic Similarity
Checker just dereferences the \texttt{output} variable to access the \texttt{Vec}
object and then accesses the symbolic contents pointed to by
the \texttt{ptr}
field pointer. These symbolic contents are 
compared against those of the \texttt{output} variable in the 
original C program.



\section{Implementation}

\sysname uses Clang 14 and Rustc v1.64 for compiling the C and
transpiled Rust code. It interfaces with the LLM models using Python modules.
Both Clang and Rustc uses the LLVM compiler
to compile C and Rust code to 
machine code, respectively. \sysname leverages this
common compiler toolchain
to compile
both the C and the Rust code to LLVM Intermediate Representation (IR), 
facilitating the subsequent symbolic execution.
\sysname uses a sequence of \emph{Clang} tools, based
on ASTMatchers~\cite{astmatchers}, to statically analyze and 
perform various preprocessing tasks on the C source
code files, such as expanding all header files, removing
unused \texttt{typedef} definitions, variable and
\texttt{struct} definitions.
%

\sysname uses the KLEE v3.1~\cite{cadar2008klee} symbolic 
execution
engine to 
symbolically execute the LLVM IR code for both the C and Rust code.
While KLEE supports C standard library functions, 
it lacks support for the Rust standard library.
Therefore, we first compiled Rust's standard library
to obtain its LLVM IR bitcode representation, and then
linked it with the bitcode of each function being symbolically
executed.
Finally, we 
developed an Antlr~\cite{antlr} parser for KLEE's \emph{Kquery} grammar
and parsed the symbolic return values to produce symbolic graphs
represented in the \emph{GraphViz} DOT
format~\cite{graphviz}.
We used the \emph{networkx}~\cite{networkx} library's
graph edit distance algorithm~\cite{hagberg2008exploring}
function to compute the edit distance between the symbolic
return values and obtain their $S^3$ score.
\section{Evaluation}
\label{sec:eval}

We evaluated \sysname with \NUMAPPS C codebases---\emph{libcsv}, 
\emph{urlparser}, \emph{optipng}, \emph{libbmp}, and \emph{u8c},
with 405, 435, 3039, 287, and 334 lines of code, respectively.
Our dataset consists of applications commonly used in previous 
works~\cite{emre2021translating,shiraishi2024context,zhang2023ownership, yang2024vert, emre2023aliasing, gao2025pr2}.
We selected the following LLM models---OpenAI's GPT-4o, GPT-4o-mini, GPT-3.5-turbo, and 
Anthropic's Claude-3.5-Sonnet, for the evaluation.

\begin{table*}[h]
    \centering
        \caption{Compilation and semantic equivalence statistics.
        Eq. Output indicates return values with $S^3$ score = 0.
        }
        
    \scalebox{0.9}{
        \begin{tabular}{l|l|rr|r|r|rr|r|r}
            \toprule
            \textbf{Codebase} & \textbf{LLM Model} & \multicolumn{3}{|c|}{\textbf{Functions}} & \multicolumn{1}{|c|}{\textbf{Code}} & \multicolumn{3}{c}{\textbf{Return values and output parameters}} & \\
             & & \textbf{C functions} & \textbf{Rust compiled} & \textbf{Perc. compiled} & \textbf{Perc. unsafe} &  \textbf{Output values}	 & \textbf{Eq. Output} & \textbf{\% of Eq. Output}	& \textbf{Bug} \\
            \midrule
libcsv      & GPT-4o            & 29 & 27  & 93.1\%  &3\%    & 73    & 62  &  84.9\%     & 5 \\
            & GPT-4o-mini       & 29 & 23  & 79.3\%  &16\%    & 66    & 55  &  83.3\%     & 1  \\
            & GPT-3.5-turbo     & 29 & 25  & 86.2\%  &15\%    & 59    & 35  &  59.3\%     & 2  \\
            & Claude-3.5-Sonnet & 29 & 15  & 51.7\%  &79\%    & 48    & 35  &  72.9\%     & 2  \\
urlparser  & GPT-4o            & 26 & 23  & 88.5\%   &10\%   & 32    & 18  &  56.3\%     & 1  \\
            & GPT-4o-mini       & 26 & 24  & 92.3\%  &27\%   & 35    & 20  &  57.1\%     & 0  \\
            & GPT-3.5-turbo     & 26 & 15  & 57.7\%  &77\%    & 24    & 20  &  83.3\%     & 0  \\
            & Claude-3.5-Sonnet & 26 & 9   & 34.6\%  &66\%    & 15    & 13  &  86.7\%     & 0  \\
optipng     & GPT-4o            & 86 & 75  & 87.2\%  &15\%    & 162   & 108 &  66.7\%     & 0  \\
            & GPT-4o-mini       & 86 & 69  & 80.2\%  &18\%    & 145   & 105 &  72.4\%     & 1  \\
            & GPT-3.5-turbo    & 86 & 59  & 68.6\%   &28\%   & 121   & 79  &  65.3\%     & 0  \\
            & Claude-3.5-Sonnet & 86 & 35  & 40.7\%  &61\%    & 69    & 54  &  78.3\%     & 1  \\
libbmp      & GPT-4o            & 16 & 16  & 100.0\% &4\%    & 38    & 35  &  92.1\%     & 0  \\
            & GPT-4o-mini       & 16 & 14  & 87.5\%  &10\%    & 42    & 36  &  85.7\%    & 0  \\
            & GPT-3.5-turbo     & 16 & 13  & 81.3\%  &12\%    & 36    & 34  &  94.4\%     & 0  \\
            & Claude-3.5-Sonnet & 16 & 13  & 81.3\%  &5\%    & 38    & 35  &  92.1\%    & 0  \\
u8c         & GPT-4o            & 19 & 17  & 89.5\%  &20\%    & 23    & 14  &  60.9\%    & 7  \\
            & GPT-4o-mini       & 19 & 16  & 84.2\%  &31\%    & 22    & 14  &  63.6\%    & 4  \\
            & GPT-3.5-turbo    & 19 & 12  & 63.2\%   &23\%   & 13    & 12  &  92.3\%    & 0  \\
            & Claude-3.5-Sonnet & 19 & 13  & 68.4\%  &57\%    & 17    & 13  &  76.5\%    & 3  \\
            \bottomrule
        \end{tabular}
    }
    \label{tab:compilation-semantic-equivalence-stats}
\end{table*}

\subsection{Compilation Results} \label{sec:Compilation}

\autoref{tab:compilation-semantic-equivalence-stats}
presents the compilation results for the \NUMAPPS codebases for the \NUMMODELS LLM models. GPT-4o outperformed all other LLM models
across all codebases, 
successfully transpiling \AVGCOMPILESUCCESSGPTFOUR of all C functions into
compilable Rust functions. GPT-4o-mini produced compilable transpilations
for \AVGCOMPILESUCCESSGPTMINI of C functions, while the GPT-3.5-turbo and
Claude-3.5-Sonnet models produced compilable Rust translations for 
\AVGCOMPILESUCCESSGPTTHREEFIVE and \AVGCOMPILESUCCESSCLAUDE of all C functions.

\para{Unsafe Rust.}
\autoref{tab:compilation-semantic-equivalence-stats} also presents the percentage of unsafe code across all \NUMAPPS codebases. 
The average percentage of unsafe code 
across all codebases is \AVGUNSAFEGPTFOUR, \AVGUNSAFEGPTMINI,
\AVGUNSAFEGPTTHREEFIVE, and \AVGUNSAFECLAUDE, for the GPT-4o, GPT-4o-mini, 
GPT-3.5-turbo, and Claude-3.5-Sonnet models, respectively. 
Similar to the compilation success results,
GPT-4o generates transpiled Rust code with the least number of unsafe program
statements. 
The reduced percentage of Unsafe Rust code for the GPT-4o 
model derives from its superior
\emph{idiomatic} type mapping capabilities.


\para{Impact of LLM model.}
\begin{figure}
    \centering
    \includegraphics[width=0.4626\textwidth]{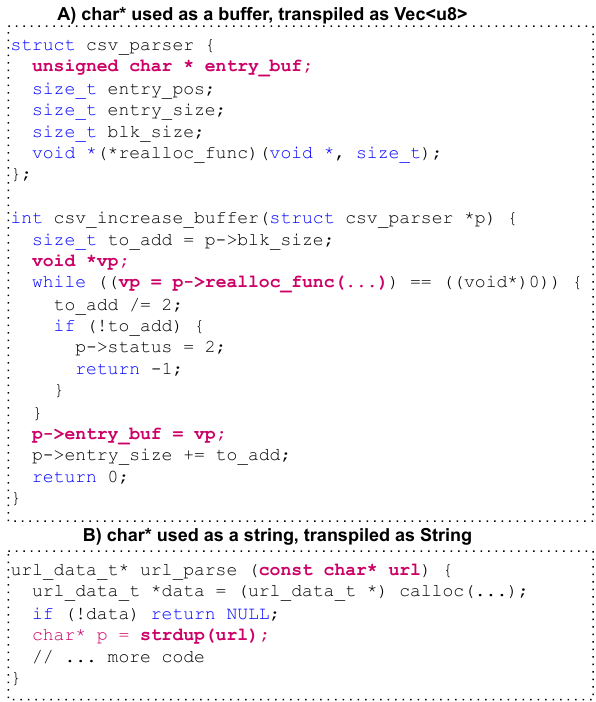} 
    \caption{Type usage based transpilation. The GPT-4o model can
    correctly identify that the \texttt{Vec<u8>} type is more appropriate
    in the first case, and the \texttt{String} type is correct for the second
    case.}
    \label{fig:type-usage-transpilation}
    \vspace{-2em}
\end{figure}
Our results show that some LLM models are more effective in
deriving 
idiomatic Rust types from C type usages.
For example, the \texttt{char*}
type variables can either be used as a null-terminated
string, or as an array of \texttt{char} type. 
Whether or not the variable should be translated to 
a Rust \texttt{string} type or a \texttt{Vec} type
depends on how it is used in the application.
Consider the simplified code snippet from 
the \emph{libcsv} codebase shown in 
\autoref{fig:type-usage-transpilation}. 
The \texttt{struct csv\_parser} 
contains a field \texttt{unsigned char* entry\_buf}. 
The \emph{libcsv} codebase uses this field 
as a char array. This is evidenced by the
presence of functions such as \texttt{csv\_increase\_buffer}
which
increases the buffer size dynamically by using the
\texttt{realloc} function, without considering any null-termination
logic. 
The GPT-4o model can correctly analyze 
the uses of this struct field, provided by the \emph{Struct Field Usage
Analyzer} described in \autoref{sec:llminvoker}, and transpile it to use the \texttt{Vec<u8>}
type. However, the other models 
simply translate
this type as a raw pointer of type \texttt{*mut u8}. 

Similarly,
the \emph{urlparser} codebase stores the URL string in a \texttt{char*} variable. 
The GPT-4o model can correctly determine that the type of the \texttt{url} argument
should be \texttt{mut String}. On the other hand, the GPT-3.5-turbo and 
Claude-3.5-Sonnet LLM models transpile this argument
as a raw pointer of type \texttt{*const c\_char}
In fact, across all \NUMAPPS codebases,
GPT-4o model generates translated Rust variables of the idiomatic \texttt{String} type
for a total of
nine times, whereas the transpilations generated by the GPT-3.5-turbo and Claude-3.5-Sonnet models generate
only one and zero variables of type \texttt{String} 
respectively, using raw pointers for the remaining variables.

\begin{table}[t]
    \caption{Rust Instruction Coverage (for compilable code)
    } 
    \
    \centering
    \scalebox{0.9}{
        \begin{tabular}{lccccc}
            \toprule
            \textbf{Codebase} & \textbf{Original} & \textbf{GPT-4o} & \textbf{GPT-4o-} & \textbf{GPT-3.5-} & \textbf{Claude-3.5-} \\ 
            & \textbf{C code} & & \textbf{mini} & \textbf{turbo} & \textbf{Sonnet}  \\
            \midrule
            libcsv &94.59\% & 90.63\% & 88.89\% & 93.96\% & 89.92\% \\ 
            urlparser &82.01\% & 72.81\% & 76.94\% & 72.21\%  & 94.02\% \\ 
            optipng &79.24\% & 78.08\% & 79.19\% & 76.51\% & 80.59\% \\ 
            libbmp &83.59\% & 79.21\% & 88.08\% & 91.77\% & 75.07\% \\ 
            u8c &95.04\% & 89.46\% & 90.19\% & 94.62\%  & 95.91\% \\ 
            \bottomrule
            \end{tabular}
    }
    \label{tab:coverage_table}
\end{table}


\begin{table}[h]
    \caption{Symbolic execution time using 8 cores (For compilable code).}
    \centering
    \resizebox{0.5\textwidth}{!}{
        \begin{tabular}{lrrrr}
            \toprule
            \textbf{Codebase} & \textbf{GPT-4o} &\textbf{GPT-4o-mini} &\textbf{GPT-3.5-turbo} & \textbf{Claude-3.5-Sonnet} \\ 
            \midrule
            libcsv       & 5h2m   & 5h2m   & 5h2m   & 5h3m   \\ 
            urlparser  & 5h4m   & 5h4m   & 5h1m   & 5h2m   \\ 
            optipng      & 15h5m  & 11h15m & 7h20m & 7h42m  \\ 
            libbmp       & 6h14m  & 3h46m  & 1h33m  & 2h0m   \\ 
            u8c          & 11h56m & 5h4m  & 5h39m  & 6h55m  \\ 
            \bottomrule
        \end{tabular}
    }
    \label{tab:execution_time}
\end{table}

\subsection{Semantic Similarity Score} \label{ref:eval_semantic_score}
We symbolized all arguments for each C and Rust
function and symbolically executed them, with a timeout of \SYMBOLICEXECTIMEOUT. 
\autoref{tab:compilation-semantic-equivalence-stats} shows the $S^3$ Score
for the C and Rust return values. Across all return values
for all applications, the 
LLM models---GPT-4o, GPT-4o-mini, GPT-3.5-turbo, and Claude-3.5-Sonnet, 
produced semantically equivalent Rust transpilations for \AVGSCOREGPTFOUR, \AVGSCOREGPTMINI,
\AVGSCOREGPTTHREEFIVE, and \AVGSCORECLAUDE of all cases.
As we discussed in \ref{sec:Compilation},
although Claude-3.5-Sonnet has the
highest 
semantic similarity score, it can only compile
\AVGCOMPILESUCCESSCLAUDE functions 
successfully, while GPT-4o can
compile \AVGCOMPILESUCCESSGPTFOUR functions. Considering the number of the 
functions that have been successfully 
compiled, GPT-4o 
provides the highest number of 
semantically equivalent 
transpilations.

\autoref{tab:coverage_table} 
shows the symbolic execution instruction coverage.
The instruction coverage for the C codebases ranges
from 79.24\% to 95.04\%, while that of the Rust code ranges 
from 72.21\% to 95.91\%.
This indicates a significant percentage of code was tested for both C program and Rust program. 
\autoref{tab:execution_time} shows
the total time taken
to execute all functions in each of the codebases 
symbolically using 8 cores. 
The Rust code symbolic execution for the GPT-4o model 
took the longest time because GPT-4o generated
the highest number of compilable functions that could
be symbolically executed.



\begin{figure*}
    \centering
    \includegraphics[width=\textwidth]{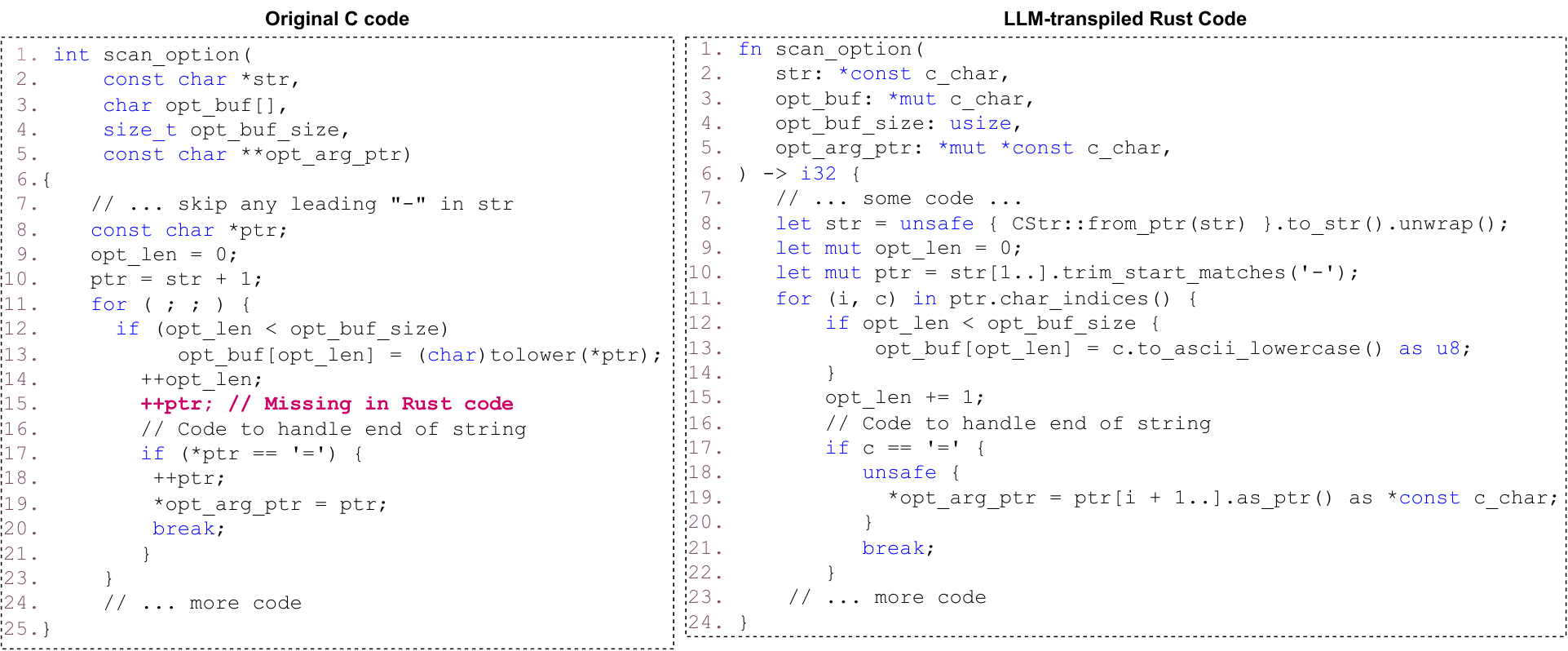} 
    \caption{Bug detected by \sysname. Rust code generated
    by Claude-3.5-Sonnet misses a pointer
    increment that results in an incorrect output.}
    \label{fig:bug-case-scan-option}
\end{figure*}


\begin{table*}[t]

  \caption{Details of bugs detected by \sysname. Note that Claude-3.5 indicates the Claude-3.5-Sonnet LLM model}
  \centering
  \begin{tabularx}{\textwidth}{@{} l >{\raggedright\arraybackslash}p{7em} >{\raggedright\arraybackslash}p{8em} >{\raggedright\arraybackslash}X @{}} 
    \toprule
    \textbf{Codebase} & \textbf{LLM Model} & \textbf{Function Name} & \textbf{Bug Description}\\
    \midrule

    libcsv &GPT-4o, GPT-3.5-turbo, GPT-4o-mini &csv\_write2 &The C version 
    writes the source 
    string into a \texttt{char*} pointer, and the Rust version writes to a string 
    slice \texttt{\&str}. If the slice is empty, indicating zero 
    capacity, the Rust code returns \texttt{0},
    whereas the C version returns the length of the input string if the output buffer is \texttt{NULL}.\\
    \hline

    libcsv &GPT-4o & csv\_set\_blk\_size &
    The Rust code has an incorrect conditional check. We discuss this in \autoref{bug:incorrect-conditional-check}. \\ \hline

    libcsv &GPT-4o & csv\_set\_delim &
    This function sets the \texttt{delim\_char} field into the \texttt{csv\_parser} struct. 
    The Rust code has an incorrect check where it skips assigning the input
    value to the field if the current \texttt{delim\_char} field is 
    ``\texttt{\textbackslash0}''. The original C code simply overwrites the 
    existing character even if it is ``\texttt{\textbackslash0}''. \\ \hline

    libcsv &GPT-3.5-turbo & csv\_free &
    This function frees the \texttt{char *} 
    field \texttt{entry\_buf} inside \texttt{csv\_parser} and sets 
    \texttt{entry\_size} to 0. Rust code uses a raw pointer for 
    \texttt{entry\_buf} and contains an incorrect check where it sets 
    \texttt{entry\_size} to \texttt{0} only if \texttt{entry\_buf} is not 
    \texttt{NULL}, whereas the 
    original C code unconditionally sets the \texttt{entry\_size} variable 
    to \texttt{0}.
    \\ \hline

    libcsv &GPT-4o, Claude-3.5 &csv\_fini & This function 
    finalizes the CSV parsing by flushing the last field. The bug
    arises because in C \texttt{switch} statements \emph{fall through}, but in 
    Rust the \texttt{match} statement does not. The LLM 
    converts a \texttt{switch} statement to a \texttt{match} statement, without accounting for the fall-through behavior. 
    \\ \hline
    
    urlparser &GPT-4o &scan\_decimal\_number &This
    function scans a string 
    and returns a pointer to the first non-digit character. The Rust 
    code returns the digit part instead of the non-digit part.
    \\ \hline

    
    optipng &Claude-3.5, GPT-4o-mini & scan\_option&
    The Rust code has a pointer increment bug. We discuss it in detail in \autoref{sec:scan_option_bug}.\\ \hline

    u8c & GPT-4o & u8encode\_ &
    This function converts a single Unicode code-point (ch) into its UTF-8 byte sequence. 
    The generated Rust code does not handle invalid Unicode sequences at all and ultimately 
    tries to prematurely add the terminator character at offset \texttt{-1} of the output 
    string slice, leading to a panic. \\
    \hline

    u8c &GPT-4o, GPT-4o-mini & u8next\_ &
    The generated Rust code returns an incorrect value when the input 
    contains invalid Unicode sequences. We discuss this bug in detail in 
    \autoref{sec:u8_next}. \\
    \hline

    u8c &GPT-4o-mini & u8next\_FAST &
    The original C function traverses the raw UTF-8 byte stream one byte
    at a time and applies bitmasks to individual \emph{bytes} to 
    reconstruct the correct Unicode code-point. The generated Rust code
    applies the bitmasks on the individual codepoints in the
    \emph{already-decoded} Unicode sequence, instead of individual bytes,
    generating garbage values. 
    \\ \hline

    u8c &Claude-3.5 & u8next\_FAST &
    When the input string is empty, the C code returns \texttt{0} in the
    output parameter \texttt{ch*}, but Rust does not update it at all, 
    leaving
    any garbage value in it intact.
    \\ \hline
    u8c &GPT-4o, Claude-3.5 & u8strncpy & This function is 
    a \texttt{UTF-8} version of strncpy that copies at most n bytes 
    from input to output and then trims any incomplete multibyte 
    sequence at the end so the output will always end with a valid UTF-
    8 character. The Rust code sets the last 
    byte of the buffer to
    \texttt{0} incorrectly, which will overwrite the meaningful 
    byte of the input.  \\ \hline
    u8c &GPT-4o, GPT-4o-mini & u8strncat &This 
    function appends characters from the source to the end of the 
    target input string and trims any incomplete multibyte \texttt{UTF-8} sequence. The Rust code performs the copy operation without 
    trimming the incomplete \texttt{UTF-8} sequence.  \\ 
    
    \bottomrule
    \end{tabularx}
  \label{tab:llm_bug_summary}
\end{table*}

\subsection{Selected Bugs}

\autoref{tab:llm_bug_summary} shows the
semantic bugs found by \sysname. 
We exclude trivial bugs, such as functions generated with
empty bodies, and
focus only on the semantic bugs.
In this section, we will discuss two such semantic bugs.

\para{Incorrect pointer increment.} \label{sec:scan_option_bug}
\autoref{fig:bug-case-scan-option} shows the simplified 
snippet of the \texttt{scan\_option} function
from the \emph{libcsv} codebase. This function parses
configuration options of the form \texttt{-option=value}.
The option name can be preceded by an arbitrary number of 
``-'' characters. In the C code, the \emph{for} loop in lines 
11-23 first scans the option name (lines 12-15). 
After the execution of line 15, \texttt{ptr} points to the \emph{next}
character to be consumed. Then, in line 17, if 
it encounters a \texttt{=} character indicating the 
separation between the option name and the value, the C
code increments the \texttt{ptr} pointer to consume and
discard the
\texttt{'='} symbol, and then stores the value to the 
\texttt{opt\_arg\_ptr} pointer. However, in the Rust
code generated by GPT-4o-mini and Claude-3.5-Sonnet, 
the next character is consumed at the \emph{end} of the
iteration. Specifically, line 15 is missing in the Rust code,
which results in the \texttt{=} symbol being appended to the
option name, instead of being discarded. Therefore, for 
an input in the form \texttt{option=value}, Rust will return
\texttt{option=} as the option name, indicating a bug.
Symbolically analyzing the original C function and the
Rust function shows that 
the minimum symbolic return value is 
\texttt{ptr = ptr + 3} for the C function
and \texttt{ptr = ptr + 2}. This shows that there is a 
bug in the Rust function and identifies that a pointer
increment operation is missing.

%
%
%

\para{Incorrect conditional check.}
\label{bug:incorrect-conditional-check}
A bug in the Rust code generated by the  GPT-4o, 
for the \emph{libcsv} codebase,
is shown in \autoref{fig:bug-case-set-blk-size}.
The original C code sets the \texttt{blk\_size} field 
in the \texttt{csv\_parser}, as long as the pointer \texttt{p}
is not \texttt{NULL}. The LLM-generated Rust code sets
the \texttt{blk\_size} field only if the previous value
of that field was not \texttt{0}. Symbolic execution
captures this semantic divergence because of the 
additional symbolic constraint \texttt{p.blk\_size != 0}.

\begin{figure}
    \centering
    \includegraphics[width=0.48\textwidth]{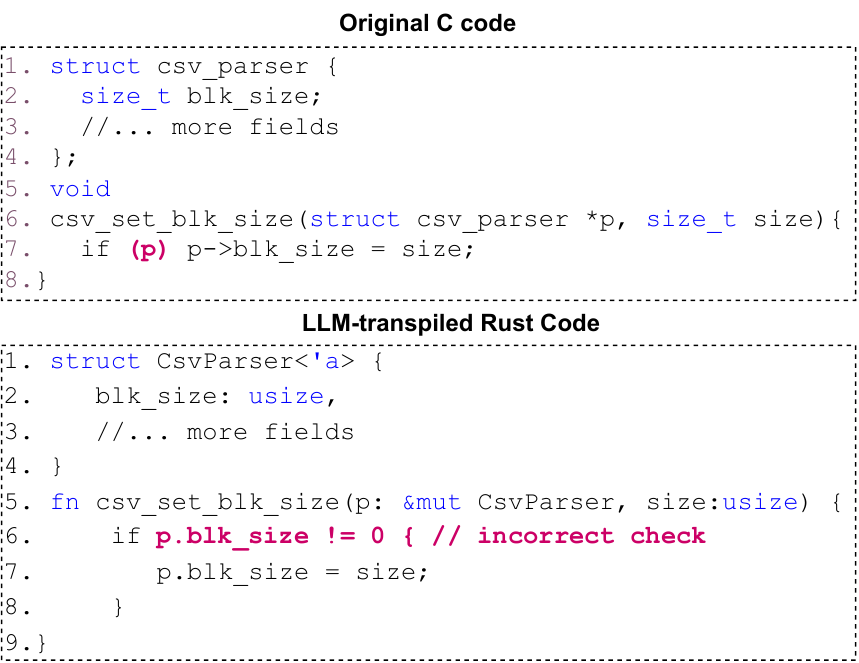} 
    \caption{Bug case detected by \sysname. Rust code 
    generated by GPT-4o has a wrong conditional check 
    that
    results in a failed assignment.
    } 
    \label{fig:bug-case-set-blk-size}
    \vspace{-1em}
\end{figure}

\subsection{Precision and Recall}

\autoref{tab:precision_recalls} shows the precision and recall analysis for 
the S$^3$ score. 
Since the recall computation requires manual verification of the 
LLM-generated Rust code, we report these statistics only for
\emph{libcsv} and \emph{u8c.}

\para{Precision.} 
The overall precision across all LLM models for \emph{libcsv} and \emph{u8c} is 85.7\% 
and 88.2\%.
Besides the semantic bugs we report in 
\autoref{tab:compilation-semantic-equivalence-stats}, the true positive 
category also 
includes LLM-generated functions that are trivially incorrect,
such as empty functions. 
The false positives arise due to optimizations
KLEE performs when symbolically modeling arrays
and aligning these differences is primarily an engineering effort.\tpalit{check}

\para{Recall.} 
To compute the recall of the S$^3$ score, we first 
determine if our system has
any false negatives with respect to the test cases contained in the 
original C codebase.
We manually
rewrote the existing test cases for \emph{libcsv} and \emph{u8c},
and used them to test the LLM-transpiled Rust code.
This includes 39 tests for \emph{libcsv}, and 5 tests
consisting of 13 test inputs for \emph{u8c}. 
Our results showed that the existing test cases could not find any new bugs
that \sysname missed. 
Only the bug in the \texttt{u8next\_FAST} code 
generated by GPT-4o-mini, that affects \emph{all inputs} 
can be detected 
by the original end-to-end test cases.

We also manually reviewed the generated
Rust code for these two codebases and did not find any bugs
missed
by our system.
Finally, as we discuss in \autoref{sec:FLUORINE}, we also use 
FLUORINE\cite{eniser2024towards} to test the transplilation results. 
Our experiments showed that, due to various crashes,
FLUORINE does not find
any bugs that cannot be revealed by \sysname.  

\begin{table}[h]
    \caption{Precision and recall analysis.}
    \centering
    \resizebox{0.5\textwidth}{!}{
        \begin{tabular}{l|l|rrrrrr}
        \toprule
        \textbf{Codebase} & \textbf{Model} & \textbf{TP} & \textbf{FP} & \textbf{FN} & \textbf{Precision} & \textbf{Recall} \\
        \midrule
        \textbf{libcsv} & GPT-4o & 5  & 2 & 0 & 71\% & 100\% \\
            & GPT-4o-mini & 7  & 1 & 0 & 88\% & 100\% \\
            & GPT-3.5-turbo      & 19 & 0 & 0 & 100\%   & 100\% \\
            & Claude-3.5-Sonnet & 5  & 3 & 0 & 63\%    & 100\% \\
        \hline
        \textbf{u8c} & GPT-4o    & 7 & 1 & 0 & 88\% & 100\% \\
            & GPT-4o-mini        & 4 & 1 & 0 & 80\%    & 100\% \\
            & GPT-3.5-turbo      & 0 & 0 & 0 &  -       & 100\% \\
            & Claude-3.5-Sonnet  & 4 & 0 & 0 & 100\%   & 100\% \\
        \toprule
        \end{tabular} 
    }
    \label{tab:precision_recalls}
    \vspace{-1.5em}
\end{table}

\subsection{End-to-end Comparison against CROWN}
\tpalit{ can we put a one line conclusion about the CROWN vs. RustAssure (other models) in the paper? Do all the models have fewer raw pointers or something like that?
We should also probably say that we compare the raw pointers because CROWN provides scripts to do that and they do not remove the unsafe keyword from functions}\gabe{Yes, all the models have fewer raw pointers than CROWN.}
CROWN~\cite{zhang2023ownership} is a  
syntax-driven transpilation toolchain that performs ownership
analysis to promote raw pointers to safe owned pointers.
\autoref{tab:compare_CROWN} compares the translation results 
of \sysname, using the best performing LLM model---GPT-4o, to those of 
CROWN.
The overall compilation success rate for \sysname and CROWN is
89.8\% and 97.3\%, respectively. 
Since CROWN is based on syntactic translation, it has 
a higher compilation success rate than \sysname, 
and does not have any runtime functional divergences,
but
it does not generate idiomatic Rust code. 
Across all codebases, \sysname and  CROWN have 35 and 271 raw pointer 
declarations, and 72 and 258 raw pointer uses, respectively, 
showing that \sysname provides more idiomatic Rust transpilations. 
Note that 
\texttt{opti\_png} declares
\texttt{18} raw pointers of which \texttt{0} are used within
the codebase---the raw pointers are strictly passed to dependent 
libraries. 
Even for the remaining models, the average raw pointer declaration and use,
normalized to the total number of Rust pointers,
is less than that of CROWN.
 
\begin{table}[h]
    \caption{Comparison of \sysname (using GPT-4o) vs CROWN.}
    \centering
    \normalsize
    \resizebox{0.5\textwidth}{!}{
        \begin{tabular}{l|rr|rr|rr}
            \toprule
            \textbf{Codebase} & \multicolumn{2}{c}{\textbf{Perc. compiled}} & \multicolumn{2}{c}{\textbf{Raw Pointer Decls.}} & \multicolumn{2}{c}{\textbf{Raw Pointer Uses}} \\
            & \textbf{Rust} &\textbf{CROWN} &\textbf{Rust} &\textbf{CROWN} &\textbf{Rust} &\textbf{CROWN} \\
            & \textbf{Assure} & & \textbf{Assure} & & \textbf{Assure} & \\
            \midrule
            libcsv     & 93\%  & 100\%  & 8  & 23  & 2  & 13   \\ 
            urlparser  & 88\%  & 88\%   & 5  & 69  & 1  & 90  \\ 
            optipng    & 87\%  & 98\%   & 18 & 135 & 0  & 66  \\ 
            libbmp     & 100\% & 100\%  & 0  & 18  & 0  & 37   \\ 
            u8c        & 89\%  & 100\%  & 4  & 26  & 69 & 52  \\ 
            \bottomrule
        \end{tabular}
    }
    \label{tab:compare_CROWN}
\end{table}

\begin{table}[h]
    \caption{Comparison of \sysname vs FLUORINE (libopenaptx library).}
    \centering
    \normalsize
    \resizebox{0.5\textwidth}{!}{
        \begin{tabular}{l|r|r|r|r|r}
            \toprule
            \textbf{Method} & \textbf{Total} & \textbf{Compilation} & 
            \textbf{Raw} & \textbf{Passes} & 
            \textbf{Passes} \\
             \textbf & \textbf{Function} & \textbf{ Success} & 
            \textbf{Pointer} & \textbf{Fuzzing} & 
            \textbf{Sym. Testing} \\
            
            \midrule
            FLUORINE   & 31 & 31 & 0 & 10 &25  \\ 
            RustAssure & 42 & 38 & 1 & N/A  &33 \\ 
            \bottomrule
        \end{tabular}
    }
    \vspace{-1em}
    \label{tab:compare_FLUORINE}
\end{table}


\subsection{FLUORINE Comparison}\label{sec:FLUORINE}

FLUORINE~\cite{eniser2024towards} 
is an LLM-based transpiler toolchain that uses
differential fuzzing to test the transpilations.
To facilitate fuzzing across C and Rust,
the toolchain also requires a JSON data-exchange file, specifying 
critical information, such as 
the number of function arguments and their types.

\para{End-to-end Comparison.} 
FLUORINE first
extracts the individual functions and their dependencies from the
C codebase.
However, the available artifact does not provide the source code
for this step.
Instead, the artifact provides the extracted functions, along
with their dependencies, and the 
JSON-mappings for 31 (out of 42) functions  
for the \emph{libopenaptx} library,
along with their Rust transpilations generated using the GPT-4o model. 
We 
transpiled \emph{libopenaptx} 
using \sysname,
and performed the comparison against the results
provided in FLUORINE's artifact.
\autoref{tab:compare_FLUORINE} shows the results of this 
comparison. Out of the 31 compilable Rust functions
generated by FLUORINE, only 10 pass FLUORINE's differential fuzzing stage,
due to crashes in the fuzzer, while \sysname can establish
the semantic equivalence of 25 out of the 31 functions handled by FLUORINE, and 33 out of all 38 compilable functions generated by 
\sysname. 
Because we manually verified the \sysname transpilations,
we did not fuzz them. This is indicated
by the \texttt{N/A} value in \autoref{tab:compare_FLUORINE}.

\para{Effectiveness on Bugs Found by \sysname.}
\label{sec:fluorine_effectiveness_bugs}
We manually crafted the JSON data-exchange file 
for the 20 buggy 
functions generated by different LLM models,
listed in \autoref{tab:llm_bug_summary},
and 
found that
only 1 of them can be detected 
by FLUORINE, with the remaining functions failing to successfully
execute to completion.
Across the 20 
tested buggy functions, 
14 failures are caused by 
the fuzzer
not supporting the argument data type. 
For example, for \texttt{Option} type function arguments, 
FLUORINE's fuzzer supports only owned values wrapped in
\texttt{Option<T>} types, and not borrowed
references (\texttt{Option<\&T>}).
Moreover, the JSON-based argument mapper assumes that 
for \texttt{struct} arguments, the transpiled
Rust \texttt{struct} definitions will have the same 
field names as their C counterparts, while \sysname does not
have this restriction.
This demonstrates how FLUORINE requires case-by-case handling
of all types.
Ultimately, \sysname can automatically handle a significantly
a larger set of types because the symbolic execution occurs at the 
LLVM IR level, which has a much smaller set of supported types. For 
example, owned values
and references are both mapped to 
the LLVM IR
\texttt{Pointer} type.
Moreover, the automatic handling of the language level differences,
described in \autoref{sec:bridging}, facilitates the comparison
of across different types.

\begin{table*}
    \footnotesize
    \caption{Comparison of C-to-Rust translation systems, approaches, strategies, and testing methods.}
    \centering
    \renewcommand{\arraystretch}{1.3}
    \begin{tabularx}{\textwidth}{@{} l >{\raggedright\arraybackslash}p{5em} >{\raggedright\arraybackslash}X >{\raggedright\arraybackslash}p{12em} @{}} 
    \toprule
    \textbf{System Name} & \textbf{Approach} & \textbf{Detailed Strategy} & \textbf{Testing Method} \\
        \hline
        LAERTES\cite{emre2021translating}  & Syntactic Translation & Uses compiler feedback to progressively rewrite unsafe pointers to owned pointers and references. & 
        Comparison of dynamic execution traces \\
        \hline
        CROWN\cite{zhang2023ownership}  & Syntactic Translation & Uses novel \textit{ownership analysis} to derive Rust ownerships and rewrite unsafe pointers to safe 
        Rust. & Existing concrete test cases \\
        \hline
        Tymcrat\cite{hong2025type}  & LLM-based & 
        Generates multiple Rust function signature candidates for each C function and uses the translated signatures of callee functions to guide idiomatic translation of callers.
        & Manually checked \\
        \hline
        \scriptsize{C2SAFERRUST\cite{nitin2025c2saferrust}}  & LLM-based & Analyzes file-level dependencies, providing them to the LLM, 
        in an optimal order. & Existing concrete test cases \\
        \hline
        RustMap\cite{cai2025rustmap}  & LLM-based & Decomposes large C 
        projects into smaller units to simplify transpilation.
        & Existing concrete test cases \\
        \hline
        \scriptsize{Code Segmentation\cite{shiraishi2024contextawarecodesegmentationctorust}} & LLM-based & Performs code segmentation by organizing C 
        functions in topological order. & Existing concrete test cases \\
        \hline
        FLUORINE\cite{eniser2024towards}  & LLM-based & Provides translation feedback to LLM, using cross-language differential fuzzer. & Differential fuzzing \\
        \hline
        SACTOR\cite{zhou2025llm} & LLM-based & Uses C2Rust\cite{c2rust} and LLMs to first translate C code to Rust and then refines pointers using ownership information 
        from CROWN\cite{zhang2023ownership}. & Existing concrete test cases \\
        \hline
        SYZYGY\cite{shetty2024syzygydualcodetestc} & LLM-based & Uses dynamic analysis to augment LLM-based transpilation.
        & Existing concrete test cases \\
        \hline
        LAC2R\cite{sim2025largelanguagemodelpoweredagent} & LLM-based & Uses LLMs to transpile C code to Rust and also generate new test cases.	& LLM-generated test 
        cases \\
        \hline
        VERT\cite{yang2024vert} & LLM-based & Generates “oracle” Rust transpilation using rWASM. & Model checking against rWASM-based oracle. \\
        \hline
        PR2\cite{gao2025pr2} & LLM-based & Lifts raw pointers in
        C2rust~\cite{c2rust} output
        to idiomatic Rust types using LLMs. 
        & Existing concrete test cases. \\
        \hline
        \textbf{RustAssure} & \textbf{LLM-based} & \textbf{Uses preprocessing and replays already transpiled structs to improve compilation success.} & \textbf{Differential symbolic testing.} \\
        \hline
    \end{tabularx}
    \label{tab:relatedwork-compare}
\end{table*}

\section{Discussion}

\para{Use Cases.}
\sysname's testing approach is not restricted to LLM-generated code 
and can be generalized to codebases transpiled using any approach.
As more teams adopt Rust for its safety guarantees, having an automated checker to validate functional parity between legacy and rewritten components 
reduces both migration effort and operational risk.

\para{Enhancements.}
\sysname can be improved with a feedback loop from the Semantic Checker 
that returns divergences to the LLM for targeted re-transpilation.
Integrating lightweight static analysis to prioritize symbolic paths, improving support for common third-party libraries, 
and caching partial executions can improve the symbolic execution coverage. 
Additionally, reducing false positives by expanding the set of language-level normalization patterns
can improve \sysname's equivalence checking. Compositional 
symbolic execution techniques~\cite{person2008differential}\cite{godefroid2007compositional} can improve 
its scalability.
\revision{
\tpalit{I think we talk about SMT solvers in the symbolic execution
section. No need to repeat it.}
}
\section{Related Work}

\para{C-to-Rust Transpilation Techniques.}
\autoref{tab:relatedwork-compare} provides a
conceptual comparison with other key approaches. 
Various analysis-based techniques~\cite{emre2021translating, emre2023aliasing, hanliangownership, ling2022rust}
transpile C to Rust using the intermediate unsafe Rust
output of the C2Rust tool\cite{c2rust}.
These approaches 
guarantee the correctness of the translation
but can produce unidiomatic Rust code\cite{eniser2024towards}.
%
%
%
%
%
Concrat \cite{hong2023concrat} focuses on replacing C Lock API
with Rust Lock API for concurrent programs.
Hong and Ryu~\cite{hong2024don} 
focus on automatically 
replacing C pointer-based return arguments
with Rust tuples.
GenC2Rust \cite{wu2025genc2rust} 
translates C code into generic Rust 
code through static analysis.
Fromherz and Protzenko~\cite{fromherz2024compilingcsaferust} 
focus on transpiling formally-verified C code to Rust.
Various approaches
combine 
program analysis techniques, such as,
ownership analysis~\cite{zhou2025llm},
type analysis~\cite{hong2025type}, 
code-slicing~\cite{cai2025rustmap}, code-segmentation~\cite{shiraishi2024context}, 
control and data flow
analysis~\cite{nitin2025c2saferrust},
with LLM-based transpilation.
%
PR$^{2}$ 
\cite{gao2025pr2} uses LLMs to
perform peephole optimization to rewrite raw pointers 
with idiomatic Rust types. 
Oxidizer \cite{zhang2024scalablevalidatedcodetranslation} translates Go to Rust by combining translation rules with LLMs.
The benchmark suites~\cite{ou2025repositorylevelcodetranslationbenchmark, khatry2025crustbenchcomprehensivebenchmarkctosaferust}
evaluate the effectiveness of C-to-Rust transpilers.



\para{Testing and LLMs.}
Current approaches for testing LLM-generated Rust code 
use existing test cases~\cite{cai2025rustmap} and differential fuzzing~\cite{eniser2024towards}. 
%
Vert\cite{yang2024vert} uses property-based testing and bounded model-checking
by compiling both the original C code and the LLM-generated Rust code
to WASM~\cite{wasm}. Unlike \sysname, however, Vert does not handle any language level differences.
HoHyun et al.~\cite{sim2025largelanguagemodelpoweredagent} presents an LLM 
agent that performs a fuzzing-based 
equivalence test to iteratively refine
Rust code, using LLMs to generate fuzzing inputs.
Syzygy \cite{shetty2024syzygydualcodetestc} combines 
LLM-driven code and test 
generation with dynamic analysis.
Various 
works~\cite{liu2023your,nunez2024autosafecoder, chen2024chatunitest} have 
used LLMs to \emph{generate} new unit tests. These 
approaches can 
complement \sysname.

\para{Symbolic execution.}
Differential Symbolic Execution~\cite{person2008differential} 
is a technique for finding diverging behavior between
different versions of the same program. \sysname extends this concept
to a cross-language setting.
Zhang et al. \cite{10.1145/3639477.3639714} created an approach to automatically verify Rust programs with KLEE.
\section{Conclusion}
We presented \sysname---an LLM-based C-to-Rust transpiler, that
performs differential symbolic testing
on the transpiled code. \tpalit{Candidate to cut. Can
just say that performs symbolic testing.} Across \NUMAPPS C codebases,
\sysname generated compilable Rust
functions for 89.8\%
of all C functions, of which 72\% produced equivalent symbolic
return values.

\bibliographystyle{IEEEtran}
\bibliography{main}

\end{document}